\documentclass[a4paper]{article}
\usepackage{ISCSLP2026}
\usepackage{ifthen}
\usepackage{multirow}
\usepackage{amsmath}
\usepackage{url}
\newboolean{blind}
\usepackage{hyperref}
\setboolean{blind}{true} 
\title{Bridging Data, Reasoning, and Alignment: A Unified Framework for Context-Aware Instruction-Following TTS}
\name{
     {Jingbin Hu$^1$, Luyu Wang$^1$, Wenjie Tian$^1$, Kangxiang Xia$^1$, Qirui Zhan$^1$,\\ Haoyu Zhang$^1$, Yunxiang Chen$^2$, Houdun Liu$^2$, Lei Xie$^{1,\star}$, Liumeng Xue$^{3,4,\star}$\thanks{$^{*}$Corresponding author.}}
}
\address{
  $^1$Audio, Speech and Language Processing Group (ASLP@NPU),
School of Computer Science, Northwestern Polytechnical University, Xi’an, China \\
$^2$ Shenzhen Pimei Technology Co., Ltd., Guangdong, China \\
$^3$ School of Intelligence Science and Technology, Nanjing University, Suzhou, China \\
$^4$ State Key Laboratory of Novel Software Technology, Nanjing University, Nanjing, China
}

\email{
    {jingbin.hu@mail.nwpu.edu.cn, lxie@nwpu.edu.cn, lmxue@nju.edu.cn, }
}

\begin{document}

\maketitle
\begin{abstract}
The ISCSLP 2026 CoT-TTS Challenge requires TTS systems to generate Chain-of-Thought (CoT) reasoning from dialogue history before synthesizing contextually appropriate speech. While the official baseline establishes a unified architecture, it remains constrained by limited contextual comprehension, weak instruction fidelity, and suboptimal audio quality. We present a systematic optimization pipeline to address these limitations. First, we develop a data process framework that cleans raw data via FullSubNet denoising, Qwen3-ASR re-transcription, and Qwen3.5-35B-A3B-based history-CoT consistency analysis, while distilling 545K high-fidelity instruction samples using Qwen3-TTS and Seed-VC under strict quality filtration. Second, we propose a Context-Aware Direct Preference Optimization (CA-DPO) method. By employing a cascaded filtering strategy, ASR prescreening, LLM tournament ranking, and speaker similarity verification, we obtain high-confidence preference pairs that significantly enhance holistic ``Context$\rightarrow$CoT$\rightarrow$Speech'' consistency during DPO training. Third, we establish an evaluation method featuring a 500-sample test set and an LLM-as-Judge framework to independently assess reasoning and execution fidelity. Experiments demonstrate that our system significantly outperforms the baseline across all objective and subjective metrics, validating our data governance and alignment strategies. Speech samples are available~\footnote{Demo:\href{https://hujingbin1.github.io/CoT-TTS-Demo-Page/}{https://hujingbin1.github.io/CoT-TTS-Demo-Page/}}.

\end{abstract}
\vspace{-15pt}
\section{Introduction}
\vspace{-5pt}

In recent years, with the advancements in large language models (LLMs) and representation learning, Text-to-speech (TTS) technology has transitioned from traditional models such as VITS~\cite{VITS} to autoregressive TTS systems dominated by large language models, such as the CosyVoice series~\cite{cosyvoice,cosyvoice2,cosyvoice3}, IndexTTS series~\cite{indextts,indextts2}, VoxCPM series~\cite{voxcpm,voxcpm2}, Qwen3-TTS series~\cite{qwentts}, and MOSS-TTS series~\cite{moss-tts,moss-vg}, as well as non-autoregressive TTS systems based on Flow Matching, such as F5-TTS~\cite{f5-tts}, MaskGCT~\cite{maskgct}, and OmniVoice~\cite{omnivoice}.

While the aforementioned systems have significantly advanced speech naturalness and intelligibility, a prominent recent trend is the pursuit of fine-grained controllability, i.e., Instruction-Following TTS (IF-TTS). The evolution of IF-TTS can be broadly categorized into three stages. The first stage relies on explicit prompt engineering or learned latent representations~\cite{promptspeaker,promptstyle,prompttts}, limited by fixed templates and restricted attribute spaces. The second stage frames controllable generation as a cross-modal alignment problem~\cite{unistyle,flespeech,histyle}, compressing textual or acoustic cues into a single continuous latent vector. However, this entangled, low-bandwidth control signal lacks explicit factorization, making fine-grained control coarse and difficult to interpret. The third and most recent stage leverages the sequence modeling capabilities of LLMs to process free-form natural language instructions~\cite{Zhou_2024,coreteam2025mimoaudioaudiolanguagemodels,hu2026voicesculptorvoicedesigned,qwentts,ov-instruct,flexvoice,moss-vg,omnivoice,voxcpm2}. Rather than compressing style descriptions into continuous vectors, these systems incorporate high-level linguistic intents directly into the language model's conditioning context, learning the intricate relationship between textual instructions and discrete audio tokens over massive, stylistically diverse datasets, thereby achieving highly flexible control over both speaker timbre and style.

Despite the remarkable controllability achieved in this third stage, existing LLM-based IF-TTS systems predominantly operate in a single-turn, explicit-instruction paradigm, assuming that the desired speaking style is directly specified by the user. In authentic conversational settings, however, the appropriate vocal style is implicitly dictated by dialogue history, speaker motivation, and situational dynamics rather than stated explicitly. This gap motivates the transition toward Context-Aware TTS. To bridge it, the ISCSLP 2026 CoT-TTS Challenge~\cite{xue2026iscslp2026cotttschallenge} introduces a novel formulation requiring the model to first generate a Chain-of-Thought (CoT) analysis, deducing the speaker's intent, emotional state, and situational context from the dialogue history, before synthesizing the target speech. By making the pragmatic reasoning process explicit through intermediate text generation, this paradigm unifies contextual comprehension and controllable speech generation into a seamless end-to-end process, where the CoT serves simultaneously as the output of contextual reasoning and the conditioning signal for stylistic synthesis.

While the CoT-TTS framework establishes a compelling architecture, building a high-performance system within this paradigm presents three critical challenges.

\textbf{First}, context-rich training data sourced from movies exhibits strong expressiveness but suffers from severe background noise and inaccurate annotations, including transcription errors, CoT-history inconsistencies, and ambiguous semantic descriptions, which collectively degrade the baseline model's semantic comprehension and audio fidelity.

\textbf{Second}, the complexity of multi-modal conditioning inputs leads to high generation variance that SFT alone cannot resolve, often yielding inaccurate CoT reasoning, poor intelligibility, and speaker identity drift. This necessitates preference-based alignment to enforce holistic ``Context$\rightarrow$CoT$\rightarrow$Speech'' consistency.

\textbf{Third}, the absence of a public context-specific benchmark and the limited granularity of conventional metrics hinder the diagnosis of reasoning-to-speech failures, necessitating a dedicated framework to independently assess contextual reasoning (Context$\rightarrow$Instruct), instruction execution (Instruct$\rightarrow$TTS), and end-to-end consistency.

In this work, we present a systematic optimization pipeline built upon the official ISCSLP 2026 CoT-TTS Challenge baseline, addressing the aforementioned challenges through three key contributions:

\begin{enumerate}
    \item \textbf{A comprehensive data processing pipeline for Context-Aware TTS.} 
    For raw data curation, we integrate FullSubNet~\cite{Hao_2021} denoising, Qwen3-ASR~\cite{shi2026qwen3asrtechnicalreport} re-transcription, and history context and CoT consistency analysis by Qwen3.5-35B-A3B~\footnote{\url{https://qwen.ai/blog?id=qwen3.5}} to filter out semantically misaligned samples. 
    For synthetic data augmentation, we leverage Qwen3-TTS Voice Design~\cite{qwentts} and Seed-VC~\cite{liu2024zeroshotvoiceconversiondiffusion} to synthesize 545K instruction samples, refined through a multi-model quality filtration pipeline to yield a high-fidelity training subset.
    
    \item \textbf{A Context-Aware Direct Preference Optimization (CA-DPO) post-training method.} 
    Through multi-candidate rollout sampling and a cascaded filtering strategy, ASR-based prescreening, Gemini 3.1~\footnote{\url{https://ai.google.dev/gemini-api/docs/models/gemini-3.1-pro-preview?hl=zh-cn}} tournament-based ranking on Context-CoT-Speech consistency, and speaker similarity verification~\footnote{\url{https://github.com/BytedanceSpeech/seed-tts-eval}}, we obtain high-confidence preference pairs. 
    Training with these pairs via DPO~\cite{rafailov2024directpreferenceoptimizationlanguage} augmented by an SFT anchor loss significantly improves contextual reasoning accuracy, output intelligibility, and speaker consistency.

    \item \textbf{A holistic evaluation method for Context-Aware TTS.} 
    We establish a bilingual (Chinese and English) 500-sample test set and an automated LLM-as-Judge framework that independently assesses Context-to-Instruct (C2I) reasoning quality, Instruct-to-TTS (I2T) execution fidelity, and end-to-end consistency, providing fine-grained diagnostic capability beyond conventional objective metrics (WER, SIM, UTMOS~\footnote{\url{https://github.com/sarulab-speech/UTMOSv2}}).
\end{enumerate}

\vspace{-5pt}
Extensive experiments on our benchmark demonstrate that our optimized system significantly outperforms the official ISCSLP 2026 CoT-TTS Challenge Track 1 baseline across all evaluation dimensions, including objective metrics, LLM-as-Judge scores, and subjective MOS ratings, validating the effectiveness of systematic data governance and preference alignment for context-sensitive speech synthesis.

\vspace{-10pt}
\section{Method}
\vspace{-10pt}
\subsection{Overview}
\vspace{-5pt}

\begin{figure}[htbp]
  \centering
  \includegraphics[width=0.5\textwidth]{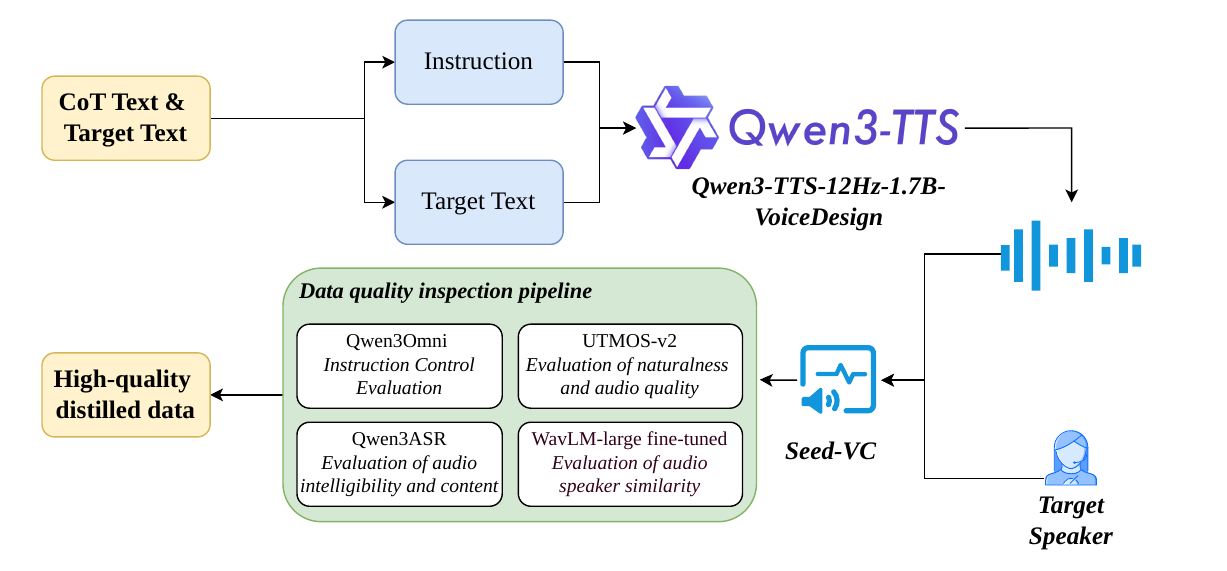}
  \caption{Synthetic Data Generation Pipeline.}
  \label{fig:pipeline}
\end{figure}

\begin{figure*}[t]
  \centering
  \includegraphics[width=0.9\textwidth]{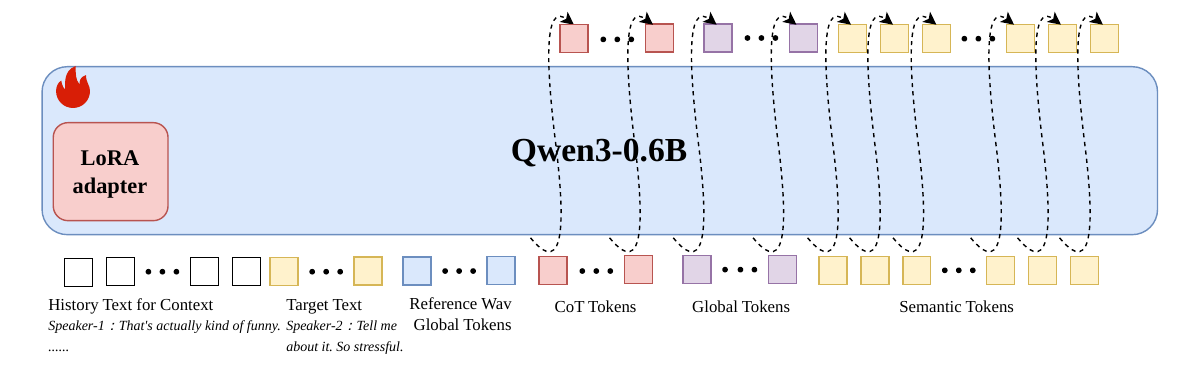}
  \caption{A Context-Aware Instruction-Following TTS Framework Based on the Official Baseline.}
  \label{fig:baseline}
\end{figure*}

\begin{figure}[htbp]
  \centering
  \includegraphics[width=0.5\textwidth]{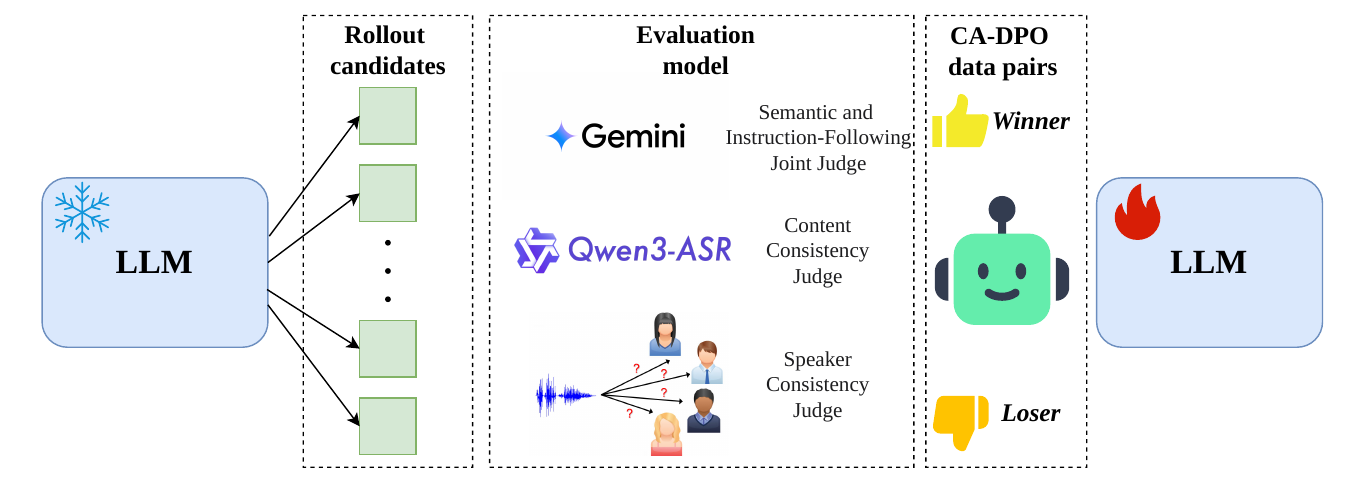}
  \caption{Context-Aware DPO.}
  \label{fig:dpo}
\end{figure}

Figure~\ref{fig:baseline} illustrates the overall architecture of this baseline system. While the baseline establishes a functional architecture, it suffers from limited contextual comprehension and suboptimal instruction fidelity due to noisy training data and the absence of preference-based alignment. We address these limitations through a systematic three-stage optimization pipeline: (i) continued pre-training on cleaned data, (ii) distillation-based supervised fine-tuning and (iii) Context-Aware preference optimization. Figure~\ref{fig:pipeline} and Figure~\ref{fig:dpo} illustrates the overall framework.

\subsection{Stage 1: Continued Pre-training with Cleaned Data}
\vspace{-5pt}
\subsubsection{Motivation}
\vspace{-5pt}

The official challenge data comprises approximately 16,000 hours of conversational speech (movie domain, English and Chinese). However, raw recordings exhibit substantial background noise, transcription inconsistencies, and variable audio quality, which degrade the performance of the base model.

\vspace{-10pt}
\subsubsection{Data Cleaning Pipeline}
\vspace{-5pt}

We design a three-stage cleaning pipeline to establish a high-quality pre-training corpus:

We design a three-stage cleaning pipeline: (i) FullSubNet~\cite{Hao_2021} denoising to mitigate acoustic degradation from complex movie recording environments; (ii) Qwen3-ASR~\cite{shi2026qwen3asrtechnicalreport} re-transcription to correct erroneous text labels and ensure reliable text-audio alignment; and (iii) Qwen3.5-35B-A3B-based consistency verification between dialogue history and CoT annotations, discarding samples with semantic contradictions or logically incoherent reasoning. Through this cascaded strategy, we progressively eliminate acoustic noise, textual errors, and semantic inconsistencies, yielding a high-fidelity corpus for continued pre-training.

Through this cascaded cleaning strategy, we progressively eliminate acoustic noise (Stage 1), textual errors (Stage 2), and semantic inconsistencies (Stage 3), yielding a stratified, high-fidelity corpus that serves as the foundation for subsequent continued pre-training.

\vspace{-10pt}
\subsubsection{Continued Pre-training}
\vspace{-5pt}

The cleaned dataset is used for continued pre-training of the baseline model. We observe that full-parameter fine-tuning significantly outperforms LoRA~\cite{hu2021loralowrankadaptationlarge} at this stage, as the model requires substantial capacity to adapt its acoustic representations to the improved data distribution. The resulting checkpoint serves as the foundation for all subsequent stages.

\vspace{-10pt}
\subsection{Stage 2: Distillation-Based Supervised Fine-Tuning}
\vspace{-5pt}

\subsubsection{Motivation}
\vspace{-5pt}

High-quality instruction-following TTS data with explicit CoT annotations is extremely scarce. To bridge this gap, we leverage knowledge distillation: a powerful teacher model generates synthetic training samples with rich stylistic control, from which the student model learns to replicate contextually appropriate speech generation.

\vspace{-10pt}
\subsubsection{Teacher Model and Data Synthesis}
\vspace{-5pt}

We employ the open-source Qwen3-TTS Voice Design model~\cite{qwentts} as the teacher. Given target text and style instructions derived from the dialogue context, the teacher synthesizes 545K speech samples with diverse emotional and prosodic characteristics. However, since the teacher model generates speech by instruction, a timbre conversion step is required to align the synthesized audio with the target speaker identity.

\vspace{-10pt}
\subsubsection{Timbre Transfer via Seed-VC}
\vspace{-5pt}

We apply Seed-VC~\cite{liu2024zeroshotvoiceconversiondiffusion} to transfer the speaker timbre of teacher-generated samples to the target speaker while preserving the rich stylistic information encoded in the original synthesis. This step ensures that the distilled training data maintains both the target speaker's acoustic identity and the teacher model's expressive control, providing the student model with high-fidelity, speaker-consistent supervision signals.

\begin{table*}[t]
\centering

\caption{Ablation study of the progressive training pipeline. We compare LoRA and full-parameter tuning at each stage, advancing the best variant. Objective (WER, SIM, UTMOS), LLM-as-Judge (1--5), and subjective scores (mean ± std) are reported, with best results in \textbf{bold}.}
\label{tab:ablation}
\resizebox{\textwidth}{!}{
\begin{tabular}{l|ccc|ccc|cccc}
\toprule
\multirow{2}{*}{Model Stage} & \multicolumn{3}{c|}{Objective Metrics} & \multicolumn{3}{c|}{Gemini Evaluation (1--5)} & \multicolumn{4}{c}{Subjective Test (1--5)} \\
\cmidrule(lr){2-4} \cmidrule(lr){5-7} \cmidrule(lr){8-11}
 & WER$\downarrow$ & SIM$\uparrow$ & UTMOS$\uparrow$ & C2I$\uparrow$ & I2T$\uparrow$ & Overall$\uparrow$ & N-MOS$\uparrow$ & S-MOS$\uparrow$ & C2I-SC$\uparrow$ & I2T-MOS$\uparrow$ \\
\midrule
Official Baseline       & 3.47 & 0.365 & 2.59 & 2.80 & 4.41 & 3.61 & 3.12$\pm$0.21 & 2.05$\pm$0.18 & 1.90$\pm$0.15 & 2.72$\pm$0.13 \\
\midrule
\multicolumn{11}{l}{\textit{Stage 1: Continued Pre-training}} \\
\quad w/ LoRA           & 3.39 & 0.364 & 2.64 & \textbf{2.87} & 4.52 & 3.70 & 3.15$\pm$0.36 & 2.08$\pm$0.14 & 1.95$\pm$0.17 & \textbf{2.90}$\pm$0.19 \\ 
\quad w/ Full           & \textbf{3.31} & \textbf{0.371} & \textbf{2.66} & 2.83 & \textbf{4.56} &\textbf{3.73} & \textbf{3.18}$\pm$0.23 & \textbf{2.12}$\pm$0.11 & \textbf{2.02}$\pm$0.08 & 2.85$\pm$0.17 \\
\midrule
\multicolumn{11}{l}{\textit{Stage 2: Distillation SFT (on best Stage 1)}} \\
\quad w/ LoRA       & \textbf{3.03} & \textbf{0.362} & 3.00 & \textbf{2.87} & 4.55 & \textbf{3.79} & \textbf{3.22}$\pm$0.28 & \textbf{2.19}$\pm$0.16 & 2.08$\pm$0.23 & \textbf{3.15}$\pm$0.14 \\
\quad w/ Full           & 3.18 & 0.357 & \textbf{3.02}& 2.86 & \textbf{4.58} & 3.74 & 3.19$\pm$0.30 & 2.01$\pm$0.29 & \textbf{2.12}$\pm$0.14 & 3.12$\pm$0.11 \\
\midrule
\multicolumn{11}{l}{\textit{Stage 3: Context-Aware DPO (on best Stage 2)}} \\
\quad CA-DPO            & 2.98 & \textbf{0.378} & 3.01 & 2.88 & 4.59 & 3.81 & 3.21$\pm$0.24 & \textbf{2.15}$\pm$0.23 & 2.02$\pm$0.16 & 3.22$\pm$0.07 \\
\quad CA-DPO + SFT anchor & \textbf{2.71} & 0.373 & \textbf{3.05} & \textbf{2.90} & \textbf{4.61} & \textbf{3.82} & \textbf{3.25}$\pm$0.19 & 2.08$\pm$0.18 & \textbf{2.14}$\pm$0.21 & \textbf{3.30}$\pm$0.20 \\
\bottomrule
\end{tabular}
}
\end{table*}

\vspace{-10pt}
\subsubsection{Multi-Model Quality Filtration}
\vspace{-5pt}

Not all teacher-generated and timbre-transferred samples are suitable for training. We apply a strict multi-model cascaded filtration pipeline:
(i) \textbf{Intelligibility}: Qwen3-ASR~\cite{shi2026qwen3asrtechnicalreport} transcribes each utterance, discarding samples with WER $\geq$ 1\%;
(ii) \textbf{Speaker Consistency}: a WavLM-based speaker verification model~\cite{Chen_2022} verifies that timbre-transferred audio maintains identity consistency with the target speaker;
(iii) \textbf{Audio Quality}: UTMOS-v2~\cite{baba2024t05voicemoschallenge2024} scores perceptual quality, removing samples with conversion artifacts;
(iv) \textbf{Style-Content Alignment}: Qwen3-Omni~\cite{xu2025qwen3omnitechnicalreport} assesses whether the acoustic realization is semantically consistent with the text content and intended style instruction.

Through this cascaded filtration, we progressively eliminate samples with intelligibility errors, speaker identity drift, acoustic degradation, and style-content inconsistencies, ultimately yielding a high-fidelity distilled training subset.

\vspace{-10pt}
\subsubsection{Supervised Fine-Tuning}
\vspace{-5pt}

We investigate both LoRA and full-parameter fine-tuning for this stage. LoRA achieves superior performance by effectively balancing the benefits of synthetic distillation data with those of authentic training data, preventing excessive adaptation to the teacher-generated distribution.

\vspace{-10pt}
\subsection{Stage 3: Context-Aware Direct Preference Optimization}

\vspace{-8pt}
\subsubsection{Motivation}
\vspace{-5pt}

The complexity of multi-modal conditioning inputs leads to high generation variance that SFT alone cannot resolve, as its token-level loss fails to distinguish globally contextually appropriate outputs from locally plausible ones. This necessitates preference-based alignment to enforce holistic ``Context$\rightarrow$CoT$\rightarrow$Speech'' consistency.

\vspace{-10pt}
\subsubsection{Multi-Candidate Rollout}
\vspace{-5pt}

For each prompt in our training set, we generate $K=10$ candidate responses through sampling, yielding diverse candidates in terms of CoT reasoning, prosodic realization, and acoustic quality. Each candidate is decoded into waveform audio for subsequent evaluation.

\vspace{-10pt}
\subsubsection{Cascaded Filtering for Preference Pair Construction}
\vspace{-5pt}

Constructing reliable preference pairs is critical for effective DPO training. We apply a three-stage cascaded filtering strategy: (i) WER pre-screening via Qwen3-ASR to eliminate unintelligible candidates; (ii) Gemini tournament ranking to identify the best candidate per group; and (iii) speaker similarity verification against the reference speaker to reject timbre-inconsistent champions.

The core of our filtering strategy is the \textbf{Gemini tournament ranking}. For each prompt group, surviving candidates participate in a pairwise ``king-of-the-hill'' tournament judged by Gemini 3.1~\footnote{\url{https://ai.google.dev/gemini-api/docs/models/gemini-3.1-pro-preview?hl=zh-cn}}. In each round, the current champion faces a new challenger, and the judge evaluates both candidates along the full ``Context$\rightarrow$CoT$\rightarrow$Speech'' chain, assessing whether the CoT reasoning correctly interprets the dialogue history and whether the synthesized speech faithfully executes the reasoning conclusions. The winner advances as the new champion until all candidates have been compared, producing a single group champion. Final preference pairs are constructed as (champion, non-champion), with losers rotated across training epochs for improved generalization.

\vspace{-10pt}
\subsubsection{DPO Training with SFT Anchor}
\vspace{-5pt}

We optimize the policy model $\pi_\theta$ by maximizing the preference reward while constraining it to remain close to the frozen reference model $\pi_{\text{ref}}$. The overall training objective combines two terms:
\begin{equation}
    \mathcal{L} = \mathcal{L}_{\text{DPO}} + \lambda \cdot \mathcal{L}_{\text{NLL}}(y_w)
\end{equation}

The DPO loss encourages the policy to assign higher probability to the preferred response $y_w$ over the dispreferred response $y_l$:
\begin{equation}
\begin{aligned}
    \mathcal{L}_{\text{DPO}} = -\mathbb{E}_{(x, y_w, y_l)} \Big[ \log \sigma \Big( \beta \big( &\log \tfrac{\pi_\theta(y_w|x)}{\pi_{\text{ref}}(y_w|x)} \\
    &- \log \tfrac{\pi_\theta(y_l|x)}{\pi_{\text{ref}}(y_l|x)} \big) \Big) \Big]
\end{aligned}
\end{equation}

Finally, to anchor the generation quality and prevent reward hacking during preference optimization, we include an SFT loss on the preferred response:
\begin{equation}
    \mathcal{L}_{\text{NLL}}(y_w) = -\mathbb{E}_{(x,y_w)} \left[ \log \pi_\theta(y_w | x) \right]
\end{equation}

\vspace{-15pt}
\section{Experiments}

\vspace{-10pt}
\subsection{Dataset}
\vspace{-5pt}

Our training pipeline utilizes three distinct datasets corresponding to each optimization stage:
\begin{itemize}
    \item \textbf{Stage 1}: The cleaned official challenge dataset, filtered by our three-stage pipeline, retaining approximately 82\% of the original samples.
    \item \textbf{Stage 2}: A high-fidelity distillation dataset refined by our multi-model cascaded filtration to yield 127K high-quality training samples.
    \item \textbf{Stage 3}: A preference dataset constructed from multi-candidate rollouts, filtered by our cascaded strategy, yielding 9K high-confidence pairs.
\end{itemize}

\vspace{-10pt}
\subsection{Evaluation Setup}
\vspace{-5pt}

We construct a bilingual (Chinese and English) 500-sample test set covering diverse dialogue scenarios. Each sample consists of a dialogue history, up to 5 turns, a target sentence, and a reference speaker audio. We report conventional objective metrics alongside human subjective scores (N-MOS, S-MOS, C2I-SC, I2T-MOS).

The core of our evaluation protocol is an automated \textbf{LLM-as-Judge} framework using Gemini 3.1, which independently assesses generation quality along three dimensions (scored 1--5):
\begin{itemize}
    \item \textbf{C2I (Context$\rightarrow$Instruct)}: Whether the CoT reasoning correctly interprets the dialogue history and derives appropriate speaking instructions.
    \item \textbf{I2T (Instruct$\rightarrow$TTS)}: Whether the synthesized speech faithfully executes the reasoning conclusions.
    \item \textbf{Overall}: End-to-end consistency of the entire ``Context$\rightarrow$CoT$\rightarrow$Speech'' chain.
\end{itemize}
\vspace{-5pt}
This multi-dimensional assessment provides fine-grained diagnostic capability for pinpointing failures in the reasoning-to-speech pipeline, complementing conventional isolated metrics. 

\vspace{-10pt}
\subsection{Main Results}
\vspace{-5pt}

Table~\ref{tab:ablation} presents the progressive ablation results across all training stages. Each stage builds upon the best-performing variant of the previous one.

We observe a clear progressive improvement from the official baseline through each optimization stage:

\textbf{Stage 1:} Full-parameter fine-tuning significantly outperforms LoRA, as the model requires substantial capacity to adapt its acoustic representations to the cleaned data distribution.

\textbf{Stage 2:} LoRA achieves superior performance over full-parameter tuning, as it effectively balances the gains from synthetic distillation data against the capabilities established on authentic data during continued pre-training. C2I and Overall scores improve substantially, indicating enhanced contextual reasoning and instruction execution.

\textbf{Stage 3:} The addition of SFT anchor loss to CA-DPO further improves all metrics. Without the anchor, pure DPO exhibits slight degradation in UTMOS and WER due to reward hacking; the anchor stabilizes generation quality while still improving contextual alignment.

\vspace{-10pt}
\section{Conclusion}
\vspace{-5pt}

In this work, we presented a systematic optimization pipeline for Context-Aware instruction-following speech synthesis, built upon the official ISCSLP 2026 CoT-TTS Challenge Track 1 baseline. We addressed three critical challenges through a unified framework: a comprehensive data processing pipeline integrating audio denoising, ASR re-transcription, LLM-based history-CoT consistency filtering, and large-scale distillation; a Context-Aware Direct Preference Optimization (CA-DPO) method that constructs high-confidence preference pairs via cascaded filtering to enforce holistic ``Context$\rightarrow$CoT$\rightarrow$Speech'' consistency; and a bilingual evaluation protocol with an LLM-as-Judge framework for fine-grained diagnosis of reasoning and execution fidelity. Extensive experiments demonstrate that each stage contributes complementary gains, with the final system significantly outperforming the official baseline across all objective and subjective metrics.

\bibliographystyle{IEEEtran}

\bibliography{mybib}


\end{document}